\documentclass[twocolumn,showpacs,prl,preprintnumbers,amsmath,amssymb]{revtex4}

\usepackage{graphicx}% Include figure files
\usepackage{dcolumn}% Align table columns on decimal point
\usepackage{bm}% bold math

\begin{document}
\title{Teaching superfluidity at the introductory level}

\author{Alexander J.M. Schmets$^{1,2}$ and Wouter Montfrooij$^1$}
\affiliation{$^1$Department of Physics and Missouri Research
Reactor, University of Missouri, Columbia, MO 65211\\
$^2$Reactor Institute Delft, Technical University of Delft, Mekelweg
15, 2629 JB Delft, the Netherlands.}
\begin{abstract}
{Standard introductory modern physics textbooks do not exactly dwell
on superfluidity in $^4$He. Typically, Bose-Einstein condensation
(BEC) is mentioned in the context of an ideal Bose gas, followed by
the statement that BEC happens in $^4$He and that the ground state
of $^4$He exhibits many interesting properties such as having zero
viscosity. Not only does this approach not explain in any way why
$^4$He becomes a superfluid, it denies students the opportunity to
learn about the far reaching consequences of energy gaps as they
develop in both superfluids and superconductors. We revisit
superfluid $^4$He by starting with Feynman's explanation of
superfluidity based on Bose statistics as opposed to BEC, and we
present exercises for the students that allow them to arrive at a
very accurate estimate of the superfluid transition temperature and
of the energy gap separating the ground state from the first excited
state. This paper represents a self-contained account of
superfluidity, which can be covered in one or two lessons in class.}
\end{abstract}
 \maketitle

\section{Introduction}
Superfluidity is the property of a liquid to flow without friction
through thin capillaries \cite{kapitza}. This property is manifest
in $^4$He below $T_{\lambda}$= 2.17 K, the so called superfluid or
lambda-transition (named after the shape of the specific heat
curve). Below 1 K, 100\% of the liquid exhibits this property.
Bose-Einstein condensation (BEC) on the other hand, is the property
that a large fraction of the particles that make up a system
condense into the same state. For instance, in an ideal Bose gas (a
gas made up of bosons that do not interact with each other), 100\%
of the particles will condense into the state with the lowest
available energy and form a Bose-Einstein condensate. However, an
ideal Bose gas does not become superfluid. And conversely, in liquid
$^4$He only about 7\% \cite{glyde} of the atoms actually form a
condensate, even though essentially 100\% of
the atoms can flow without friction below 1 K.\\

In fact, there is no reason why a system could not become a
superfluid even if only a very small fraction of the atoms were to
form a condensate. All this nicely illustrates the fact that
superfluidity and BEC are two different phenomena, even though
introductory textbooks tend to lump the two together. The main
difference between BEC and superfluidity is that BEC is a property
of the ground state, while superfluidity is a property of the
excited states. This is entirely analogous to standard
superconductivity, where the electrons condense into Cooper pairs
(ground state), and where the interaction between the Cooper pairs
introduces a finite energy gap between the ground state and excited
states. In turn, this energy gap is responsible for the system
becoming a superconductor. Thus, in both systems, it is the
interaction between the particles that is responsible for
the exotic behaviors, not how they arrange themselves in the ground state.\\

In this paper we focus on the property of superfluidity rather than
on BEC. We repeat Feynman's arguments that show that any Bose liquid
that stays liquid down to low enough temperatures must become a
superfluid because of the presence of an energy gap. We also derive
a very accurate estimate of the superfluid transition temperature
using basic conservation laws and some straightforward
approximations. Altogether, this should give students a much better
understanding of what superfluidity entails and why it necessarily
must occur in $^4$He. In addition, our simple calculations should
bestow upon them the idea that they have already learned enough
physics to be able to come up with a very accurate estimate of
something as complex as the superfluid transition temperature in
$^4$He. Note that we do not argue that a Bose condensate does not
form in $^4$He, rather we focus on the majority of non-condensed
atoms (93 \%) that determine the numerical values for virtually all
quantities of interest.\\

\section{Superfluidity: qualitative understanding}

\begin{figure}[t]
\includegraphics*[viewport=75 100 600 450,width=90mm,clip]{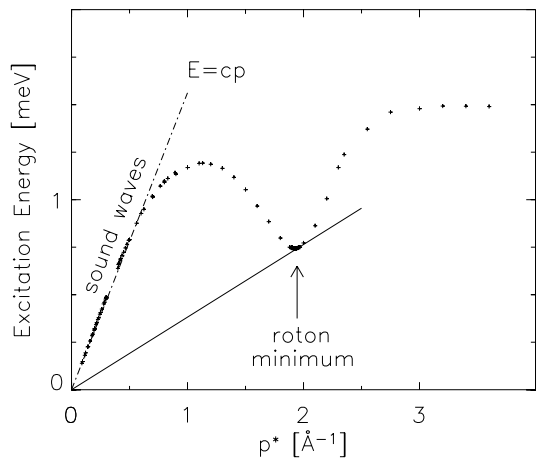}
\caption{The measured excitation energies $E(p)$ \cite{dispcurve} in
superfluid $^4$He as a function of momentum transfer $p^* = p/\hbar=
2\pi/\lambda$. The slope of the curve at small $p^*$ [dashed line]
is given by the velocity of sound $c$= 237.4 m/s \cite{sound}, the
overall minimum slope [corresponding to a speed of 58 m/s] is given
by the solid line which is tangent to the excitation curve near the
so-called roton minimum [$p^*= 1.94$ \AA$^{-1}$, $E(p)$= 0.743
meV].} \label{roton}
\end{figure}

\begin{figure} [t]
\includegraphics*[viewport=75 100 600 330,width=130mm,clip]{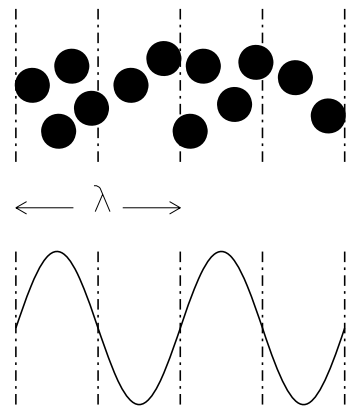}
\caption{A real space visualization of a density disturbance [a
departure from the average density of 3 in this figure] that
resembles a sound wave with a wave length $\lambda$ of about 3 times
the average atomic separation. The roton minimum corresponds to a
disturbance with a wave length comparable to the atomic separation,
but there is no agreement on how to visualize such a short wave
length excitation \cite{rotonviz}} \label{densfluc}
\end{figure}

\begin{figure} [t]
\includegraphics*[viewport=75 130 600 480,width=90mm,clip]{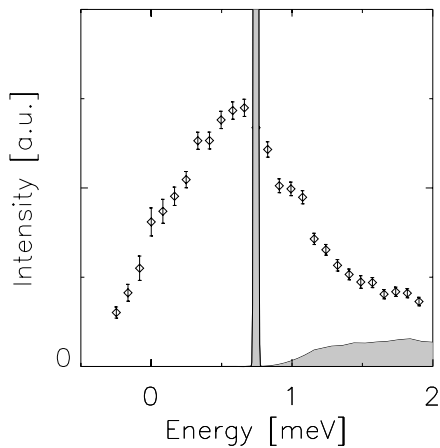}
\caption{Detailed view of the excitations of $^4$He corresponding to
the roton minimum in the superfluid phase (shaded) and in the normal
fluid phase (points plus errorbars) \cite{mont2}. The data are taken
using neutron scattering. The neutron transfers energy $E$ to the
liquid and an amount of momentum corresponding to the roton minimum
[see vertical arrow in Fig. \ref{roton}]. When the amount of energy
transferred exactly matches the energy difference between the ground
state and the excited state, then a sharp resonance peak [at 0.743
meV] can be seen in the superfluid. Note that there is no signal
below this peak. In the normal fluid the behavior is very different;
even a small amount of energy is sufficient to excite the liquid,
and there is a clear signal even at $E$ =0. The signal at $E <$ 0
implies that the liquid gives up some of its energy to the neutron,
which can happen if the liquid is not at zero Kelvin. The height of
the peak of the sharp resonance is such that the size of the shaded
area is the same as the area under the curve for the normal fluid.}
\label{sfluids}
\end{figure}

First, the fact that helium does not solidify at any temperature is
a pure quantum effect. The weak van der Waals forces between the
atoms are not strong enough to overcome the zero point motion
associated with trying to confine a helium atom to a lattice site.
The second aspect that makes helium stands out from other liquids is
that it takes a finite amount of energy to create a disturbance in
the liquid. This is shown in Fig. \ref{roton}. The actual amount of
energy required depends on the wavelength $\lambda$ (or momentum
$p=h/\lambda$) of this disturbance. The measured values
\cite{dispcurve} for the energy cost are shown in Fig. \ref{roton}.
At low momentum transfers (long wave lengths) the energy disturbance
is just a run-of-the-mill sound wave (a phonon), and its energy is
given by $E_{ex}(p) = c p$, the standard hydrodynamics result for any liquid,
not just superfluids \cite{hydro}.\\
When we go to lower wave lengths, such as the density disturbance
pictured in Fig. \ref{densfluc}, the energy cost starts to deviate
from $E_{ex}(p) = c p$. For wave lengths comparable to the
interatomic spacing $d$, the energy cost goes through a minimum,
after which it goes up again. This minimum of the energy gap between
the ground state and the excited state is commonly referred to in
the literature on superfluid helium as the roton minimum
\cite{rotonviz} or simply 'the roton', and the entire curve is
referred to as the phonon-roton dispersion curve. The roton turns
out to be the determining feature of superfluids. As pointed out in
the preceding, the presence of this roton gap is analogous to the
presence of a similar gap in superconducting systems. The presence
of a gap also firmly sets superfluid $^4$He apart from normal fluids
where nothing resembling an energy gap exists. This is shown in Fig.
\ref{sfluids} where we compare
helium in the superfluid phase to helium in the normal fluid phase.\\
One can easily verify from Fig. \ref{roton} that the presence of a
non-zero energy gap is synonymous with superfluidity. The slope of a
line that goes through the origin and a point on the excitation
curve gives the (group) velocity of the excitation. For instance,
this slope at small momenta is given by the speed of sound [see Fig.
\ref{roton}]. The overall smallest slope is encountered near the
roton minimum. The value of the slope at this point corresponds to
the velocity below which the liquid can flow without friction, at
least through small capillaries \cite{caveat}. After all, if the
liquid is flowing at a lower speed, then the liquid cannot slow down
because of the following restrictions due to the energy and momentum
conservation laws. Following standard arguments \cite{kittel}, we
focus on a liquid mass $M$ that is flowing at speed $v$. For it to
slow down to speed $v'$ by creating an excitation of energy $E_{ex}$
and momentum $\overrightarrow{p}_{ex}$ we have
\begin{equation}
\begin{array}{c}
Mv^2/2=Mv'^2/2+E_{ex} \\
M\overrightarrow{v}=M\overrightarrow{v}'+\overrightarrow{p}_{ex}.
\end{array}
\label{basic}
\end{equation}
Eliminating $v'$ we get
\begin{equation}
\overrightarrow{v}.\overrightarrow{p}_{ex}-p^2_{ex}/2M=E_{ex}.
\label{basic2}
\end{equation}
Even in the best case scenario in which $\overrightarrow{v}$ and
$\overrightarrow{p}_{ex}$ are parallel and in which $M$ is very
large we find the minimum requirement on the flow velocity $v$ for
the liquid to be able to slow down:
\begin{equation}
v \geq E_{ex}/p_{ex}
 \label{basic3}
\end{equation}
In a normal liquid without an energy gap, liquid flow will always be
damped because the minimum slope would be zero. This [Eq.
\ref{basic3}] of course is also the reason why an ideal Bose gas
does not become superfluid. Here the excitation energies are given
by $E_{ex}= p_{ex}^2/2m$, and a parabolic curve does not have a
minimum slope: no matter how slow an ideal Bose liquid is flowing,
it is always possible to transfer energy by creating an excitation,
and the liquid will slow down. Also, note that even though it
requires less energy to create a sound wave than a roton excitation,
the roton minimum actually
determines the critical flow velocity.\\
In this paper we explain why there is an energy gap in the first
place, how the size of this gap relates to the superfluid transition
temperature, and how actual values for all parameters involved can
be estimated. Feynman explained in a beautiful argument why this
energy gap is the unavoidable consequence of the fact that $^4$He
atoms obey Bose statistics. We refer the reader to Feynman's 1955
account \cite{feynman} and 1972 textbook \cite{textbook} for details,
but in a nutshell the argument is the following.\\

Assume that a certain configuration of the helium atoms represents
the state with lowest energy, the ground state. The quantum
mechanical wave function $\phi$ of this state depends on the
positions of all atoms:
$\phi(\overrightarrow{R}_1,\overrightarrow{R}_2,..\overrightarrow{R}_N)$.
The energy of this ground state consists of a kinetic energy term
that depends on the gradient of the wave function $\sim |\nabla
\phi|^2$ as well as a potential term $V|\phi|^2$. [The same holds
for the wave function
$\psi(\overrightarrow{R}_1,\overrightarrow{R}_2,..\overrightarrow{R}_N)$
describing the excited state that is lowest in energy of all excited
states]. The potential operator has terms $\sim
1/|\overrightarrow{R}_i-\overrightarrow{R}_j|^n$ which tell us that
the force between the atoms
is strongly repulsive when they are too close together.\\
\begin{figure}[t]
\includegraphics*[viewport=150 130 650 260,width=130mm,clip]{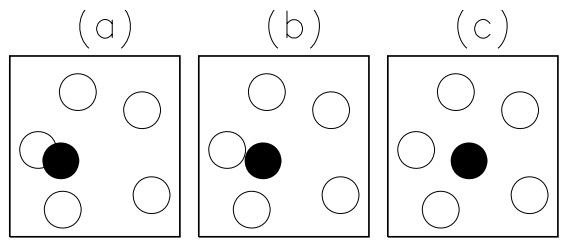}
\caption{A depiction of various configurations representing
different energies \cite{textbook}. Part (a) shows an unlikely
ground state configuration since two atoms being in the same spot
implies a high potential energy. Similarly, part (b) shows an
unlikely ground state because this configuration would represent a
high kinetic energy [see text]. On average, in the ground state the
atoms will be spaced out as shown in part (c).} \label{groundstate}
\end{figure}
The exact details of the ground state are not important, but both
the kinetic and potential term should be small. From this
requirement we can expect that the atoms in a configuration that
could represent the ground state are fairly well spread out [see
Fig. \ref{groundstate}c]. After all, if they were to sit on top of
each other [Fig. \ref{groundstate}a], we would pay a high price in
potential energy, and hence, we can assume the amplitude of the
ground state wave function $\phi$ to be zero for those cases where
$R_i \approx R_j$. Also, the atoms will not be too close to each
other [Fig. \ref{groundstate}b], because this would correspond to a
high gradient $\nabla \phi$, making it an unlikely choice of ground
state. We can see that atoms almost touching each other would
corresponding to a high gradient as follows: if the amplitude of
$\phi$ would not be zero for a configuration where two atoms are
very close, then we would have the situation that by slightly
changing the coordinate of one atom to make it sit on top its
neighbor [going from Fig. \ref{groundstate}b to \ref{groundstate}a],
we would go from a non-zero to a zero amplitude for $\phi$. This
implies a steep gradient $\nabla \phi$, and therefore, the amplitude
of $\phi$ must also be zero for configurations where atoms are too
close. Another way of saying the above is that if an atom actually
were to move from Fig. \ref{groundstate}b to \ref{groundstate}a, it
must have had a large kinetic energy in the first place to be able
to approach the other atom as closely as shown in Fig.
\ref{groundstate}a. However, note that we do not actually 'move'
atoms, we just compare the wave function for two different
configurations $\overrightarrow{R}^N$. Thus, when we say 'move', we do not imply any dynamics.\\
\begin{figure}[t]
\includegraphics*[viewport=75 130 600 430,width=90mm,clip]{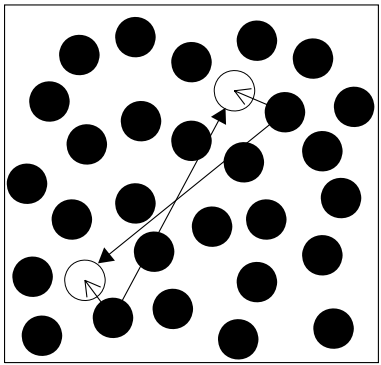}
\caption{A reasonable guess of what a low energy state in helium
could look like \cite{textbook}. To flip the sign of $\psi$ atoms
are 'moved' over large distances (long arrows) while smoothing out
any holes left in the liquid to minimize the energy cost. However,
since the Bose particles cannot be distinguished from each other and
since permutation of particles does not affect the wave function,
the outcome of the 'movements' indicated by the solid arrows are
identical to the 'movements' indicated by the short arrows.}
\label{excitedstates}
\end{figure}
We also know that $\phi$ will not have any nodes other than at the
edge of the box confining the liquid, so we can assume that the
amplitude of $\phi$ is positive for all configurations. Also, the
excited state wave function $\psi$ should have at least one more
node. This is essentially the guitar string equivalent of looking at
the first and second harmonic. This implies that half of the
configurations representing the excited state
$\psi(\overrightarrow{R}_1,\overrightarrow{R}_2,..\overrightarrow{R}_N)$
correspond to a positive amplitude, and half of the configurations
correspond to a negative amplitude \cite{scalarproduct}. We will now
try to create an excited state that barely differs in energy from
the ground state. This way, the roton minimum would be very small
[yielding a very small minimum slope] and we would be able to stifle
superfluidity. However, Feynman showed that this cannot be done
\cite{feynman}, and that one always ends up with a sizeable energy
difference between the ground state and the excited states [that are not phonons].\\

In Fig. \ref{excitedstates} we have sketched a configuration that we
arbitrarily will take to correspond to a maximum positive amplitude
for $\psi$. We do not really know what the configuration should look
like, but we tried to make it look like the ground state, with atoms
not sitting on top of each other. Next, we will rearrange the atoms
to end up with a configuration that would correspond to a maximum
negative amplitude. To achieve this, we should rearrange the helium
atoms over large distances. We are not interested in short
distances, because this would imply that $\psi$ goes from a maximum
to a minimum over short distances, which would correspond to a large
gradient $\nabla \psi$ and therefore, to a high energy. We will also
smooth out any holes or bumps that may materialize, otherwise we
would end up with an excitation that looks like a phonon (see Fig.
\ref{densfluc}) and we already know that a phonon does not represent
the minimum slope of the excitation curve. The required changes to
the configuration are shown in Fig. \ref{excitedstates}
by the long arrows, and we appear to have achieved our aim.\\

However, the above approach does not work in a Bose liquid since all
the atoms are indistinguishable and interchanging two atoms does not
lead to a change in the amplitude of the wave function. Thus, one
could have gotten the same final configuration by simply 'moving'
the atoms affected by the rearrangement over distances less than
half the atomic separation [short arrows in Fig.
\ref{excitedstates}]. In fact, half the atomic separation is the
best that one could achieve. However, such a rapid variation (from
maximum to minimum over half the atomic separation) would represent
a large gradient and signify a significant step up in energy. In
other words, because of the Bose nature of the atoms, it is not
possible to make an excited state (which is not a phonon) that
differs by a vanishingly small amount in energy from the ground
state. Therefore, an energy gap must be present in a Bose liquid and
provided the liquid does not freeze, it must become a superfluid.
Whether a Bose-Einstein condensate forms or not is not relevant to
this argument since it links the property of superfluidity to the
scarcity of excited states. Even a Bose liquid where only a tiny
fraction of the atoms condenses will become a superfluid when the
temperature is low compared to the energy gap.\\

Thus, from a qualitative point of view, it is clear why a Bose
liquid that remains liquid down to low enough temperatures has to
become a superfluid because of the presence of an energy gap.
However, this argument does not provide us with a numerical estimate
for the size of the gap. Moreover, it does not even tell us how the
transition temperature is linked to the size of a gap. This probably
explains why textbooks tend not to mention Feynman's arguments.

\section{Superfluidity: quantitative understanding}

So how low does the temperature have to be for the liquid to become
a superfluid? Since we do not think that the temperature at which
BEC occurs in an ideal Bose gas made up of non-interacting atoms
with the same mass as helium atoms [3.13 K] has much to do with the
magnitude of the energy gap in a real liquid, we must take a
different approach. We first discuss a relationship between the size
of the energy gap and the superfluid transition temperature,
followed by a discussion on how to estimate the size of the gap
based on the speed of sound and the particle density of the liquid.
Doing so, we will end up with an accurate estimate of the superfluid
transition temperature. Our estimates apply to $^4$He at zero
pressure, possible extensions to higher pressures are given in
footnotes.\\

As a note of caution, while our estimates turn out to be remarkably
accurate [probably because of cancelation of errors], the following
should not be read as anything other than being a set of
instructional exercises to help students in their understanding of
liquid helium in particular, and in using their acquired knowledge
from introductory physics to real world problems. It is not a
derivation of the transition temperature in superfluid helium,
even though our estimate turns out to be very accurate.\\
\subsection{Connection $T_{\lambda}$ to excitation gap}
We first apply the same reasoning as in the preceding section to
figure out what the minimum velocity requirement is for a single
helium atom. For this atom of mass $m$ moving with speed $v$ through
the sea of other atoms to be able to slow down to speed $v'$ we
have:
\begin{equation}
\begin{array}{c}
mv^2/2 = mv'^2/2 + E_{ex}(p)\\
m\overrightarrow{v}=m\overrightarrow{v}'+\overrightarrow{p}_{ex}.
\end{array}
\end{equation}
Since we are interested in the bare minimum, we assume that this
atom will give up all of its energy [$v'$=0]. Dividing the above
equations we get a minimum condition on the speed of the atom for it
to be able to transfer energy to the rest of the liquid:
\begin{equation}
v_{minimum} \geq 2[E_{ex}/p_{ex}]_{minimum} \label{above}
\end{equation}
Compared to Eq. \ref{basic3} we have picked up a factor of 2, which
is the result of dealing with a small mass $m$ instead of with the
large mass $M$ of a moving liquid. Note that the actual mass of the
atom does not play a role in this expression. This ensures the
validity of Eq. \ref{above} in real liquids in which the actual
movement of an atom is accompanied by a flow pattern where atoms are
temporarily pushed out of the way, bestowing the moving atom with an
effective mass which is larger than that of the mass of a
non-interacting atom
[about 2$\sim$3 times larger, see ref \cite{wouter} for details].\\

If an atom moves faster than this minimum requirement it can create
an excitation in the rest of the liquid and slow down, if it moves
slower then it will not be able to slow down (that is, it will not
experience any friction). Thus, we should expect to see a
qualitative difference in behavior of the liquid above and below the
temperature at which the thermal velocities meet the minimum
requirement contained in Eq. \ref{above}. To estimate this
temperature, we assume that the classical equipartition of energy
principle can be extended to quantum liquids at low temperature,
namely $mv_{thermal}^2/2 = 3k_B T/2$. Combining this with Eq.
\ref{above} we find that we can expect changes in liquid behavior at
a temperature $T_{\lambda}$ when $v_{thermal} = v_{minimum}$:
\begin{equation}
T_{\lambda}= \frac {4m[E_{ex}/p_{ex}]^2_{minimum}}{3k_B}.
\label{lambda}
\end{equation}

We read off the value of the minimum slope \cite{wouter} from Fig.
\ref{roton}: ${(E_{ex}/p_{ex})}_{minimum}$= (58.05 $\pm$ 0.15) m/s
which combined with the mass of a helium atom of 6.646 x 10$^{-27}$
kg yields a transition temperature of $T_{\lambda}$ = 2.162 K $\pm$
0.012 K, in good agreement with the actual transition temperature of
2.17 K. The fact that the agreement is essentially perfect might be
fortuitous, however, it does show that our assumption of being able
to use the equipartition of energy theorem to be not too far off the
mark \cite{extend}. Also, Eq. \ref{lambda} tells us that the
transition temperature is determined by an intrinsic microscopic
velocity of the liquid, as we would have expected
for superfluidity.\\
\subsection{Connection excitation gap to speed of sound}
Now that we have made the connection between the macroscopic
transition temperature and the microscopic parameters for the roton,
we can try to estimate these roton parameters based on other
macroscopic quantities. We start with the value of the energy gap at
the roton minimum. As an aside, we note that the entire excitation
curve shown in Fig. \ref{roton} can in principle be calculated with
great precision from first principles \cite{principles}, including
the flat part at higher momentum values [$p^* \geq 3$ \AA$^{-1}$]
and its termination at twice the roton energy \cite{mont2}. The
procedure is straightforward, but cumbersome. Since we only need the
value of the
energy gap at the roton minimum, we use the following more instructive shortcut.\\

As Feynman pointed out in his argument why a Bose liquid should
exhibit an energy gap \cite{feynman}, one gets the lowest lying
excited state when one 'moves' atoms by half the atomic separation.
We can use this to calculate the value of the energy gap based on
the macroscopic speed of sound in superfluid helium. From
thermodynamics we have that the speed of sound $c$ is given by
\begin{equation}
c^2= \frac{\gamma}{m} \left( \frac{\partial P}{\partial n}\right)_T
\end{equation}
with $P$ the pressure and $\gamma$ the ratio of specific heat at
constant pressure $c_p$ and at constant volume $c_v$. For superfluid
helium at low temperature we find that $\gamma =1 $ \cite{sound} as
a direct consequence of the large zero-point motion of the
atoms\cite{explain}. Thus, at constant volume $V$ we have
\begin{equation}
c^2= \frac{1}{m} \left( \frac{\partial PV}{\partial N}\right)_{T,V}.
\end{equation}
We can now calculate $c$ by adding one more atom to the liquid at
constant volume. In this case, $\triangle N$ =1 and $\triangle PV$
is the amount of work we have to do to make room for this additional
atom.\\

We calculate this amount of work by comparing the energy of a
configuration with a hole in it to the energy of a configuration
without such a hole. We can make a cubic hole in the liquid by
'moving' atoms along the positive x-direction by half the atomic
separation $d/2$, and by doing the same thing along the negative
x-direction, and by repeating the process in the y and z-directions.
Each of these 6 configurational changes should increase the energy
of our state, but only by $E_{roton}$ for every step. This can be
seen as follows: provided we can 'move' atoms over a distance of at
least $d/2$, and provided we have plenty of room to smooth out any
variations in local particle density, then we should be able to do
each step of the process at a minimum energy cost. Thus the total
cost will be six times the minimum excitation energy in liquid
helium, or 6$E_{roton}$. Of course in doing this, we actually made
the hole too big since we only needed to make a sphere of diameter
$d$. In all, we only need to provide
$6E_{roton}[(4\pi/3)(d/2)^3]/d^3= \pi E_{roton}$ in work. Combining
all this we find \cite{higherpres}
\begin{equation}
E_{roton}= m c^2/\pi. \label{rotonenergy}
\end{equation}
To see how reliable an estimate this is, we compare this prediction
to the measured quantities of superfluid helium at 1.2 K. Using c =
237.4 m/s \cite{sound} (Fig. 1), we obtain $E_{roton}$ = 11.92 x
10$^{-23}$J = 0.744 meV. The value that has actually been measured
by means of neutron scattering \cite{dispcurve} is 0.743 meV. Thus,
we have found a very accurate value for the energy gap based on the
speed of sound. In essence, we have used the speed of sound to gauge
the strength of the interatomic potential, which in turn
determines the roton energy.\\
\subsection{Connection roton excitation to liquid density}
So far we have connected the superfluid transition temperature to a
minimum speed which can be determined from the excitation curve
shown in Fig. \ref{roton}, and we have connected the minimum
excitation energy $E_{roton}$ to the macroscopic speed of sound. To
be complete, we should also estimate the momentum value
corresponding to this minimum excitation energy, so that we can get
an overall estimate of the minimum value of $E_{ex}/p_{ex}$ for the
excitations shown in Fig. \ref{roton}. We expect
$E_{roton}/p_{roton}$ to be a very accurate estimate of this
minimum, perhaps fractionally too large [see Fig. \ref{roton}]. The
roton momentum $p_{roton}$ is given by $p_{roton}=
h/\lambda_{roton}= h/d$, with $h$ Planck's constant and $d$ the
interatomic separation. In words, the energy cost to create an
excitation of wave length $\lambda$ is least when this wave length
matches the natural length scale in the liquid, the atomic
separation.\\

The reader might be surprised to see that the roton wave length
corresponds to $d$ instead of $d/2$. This goes back to our earlier
usage of the verb to 'move'. We did not actually move a single atom
over $d/2$, nor did we say that moving a single atom would
correspond to the roton excitation. The only fact we used was that
the entire configuration described by the excited state wave
function $\psi$ shown in Fig. \ref{excitedstates} contained a roton
excitation. The question we ask now is subtly different: if we were
to characterize the excited state wave function by the average
distance between atoms, what distance would we find? The answer is
$d$ since the excited state is similar to the ground state in many
ways [in fact we constructed it so that the two would be as close as
possible]. Whatever a roton excitation might be, it is some
arrangement where the average separation between the atoms is pretty
much as it is in the ground state, otherwise the energy of the
excited state would be much higher.\\

The separation $d$ depends on the number density $n$ of liquid
helium as $d \sim n^{-1/3}$. Estimating the proportionality factor
is a somewhat nebulous undertaking because unlike in a solid, we do
not have a nice periodic arrangement. For our estimate we use that
helium fairly accurately resembles a liquid of closely packed
spheres. The fraction of the volume occupied by the atoms in such a
liquid is $\pi/3\sqrt 2$= 0.741, so that we estimate the atomic
separation to be \cite{more}
\begin{equation}
d=[\pi/(3n\sqrt 2)]^{1/3}; p_{roton}=h[\pi/(3n\sqrt 2)]^{-1/3}.
\label{rotonpos}
\end{equation}
This separation is slightly lower than that for a simple cubic
structure [$d=(1/n)^{1/3}$] which simply tells us that the atoms are
closer together in a closely packed structure. Combining Eqs.
\ref{lambda}, \ref{rotonenergy} and \ref{rotonpos} we get
\begin{equation}
T_{\lambda}= \frac {4m}{3k_B}[\frac{mc^2(\pi/3\sqrt 2)^{1/3}}{h \pi
n^{1/3}}]^2.
\end{equation}

When we plug in all the numbers [c= 237.4 $\pm$ 0.5 m/s, n= 0.02183
atoms/\AA$^3$], we find $T_{\lambda}$= 2.17 $\pm$ 0.02 K. Thus, when
we combine our quantitative approximation for the relationship
between the minimum of the dispersion curve and $T_{\lambda}$ [Eq.
\ref{above}], with our approximation for the roton energy [Eq.
\ref{rotonenergy}], and with our approximation for the roton
position [Eq. \ref{rotonpos}], we still find very good agreement
with experiment. While it is satisfying to have ended up with such a
good agreement, we note that the agreement is probably better than
we had reason to expect given the simplicity of our estimates. From
an instructional point of view however, we consider the individual
links that we have made between microscopic parameters and
macroscopic quantities [that is, Eqs, \ref{above}, \ref{rotonenergy}
and \ref{rotonpos}] to be the most important points of this section
since it allows students to apply basic physics reasoning in order
to arrive at
predictions for measurable quantities.\\

In summary, we have shown that superfluidity can be explained to
students without going into lengthy calculations, and without having
to invoke a Bose condensate. We have tied Feynman's arguments about
the origin of the energy gap to the actual superfluid transition
temperature, and we have shown that very accurate estimates of all
parameters involved can be obtained through straightforward
reasoning. While we have included the actual numerical values for
our calculations in this paper to make the discussion less abstract,
and while we have even given an expression of the superfluid
transition temperature in terms of macroscopic quantities like
density and speed of sound, the real message of the paper is that
students should be able to develop a better sense of what causes the
property of superfluidity in terms of Bose statistics and energy
gaps; that is, better compared to the standard Bose-Einstein
condensation remarks that are normally encountered in textbooks.
Finally, since a very similar relationship exists between the
transition temperature and the energy gap in superconductors, this
paper could also serve as an
introduction to the physics of superconductivity.\\


\begin{thebibliography}{10}
\bibitem{hydro} The reader can verify that a linear relationship
is common to all systems by listening to a piece of music and
noticing that the high and the low tones travel just as fast, which
can only happen provided $E/p$ is constant for small $p$.
\bibitem{rotonviz} Different visualizations of what
exactly a roton is have been put forward over the years. For
instance, it can be described as a sound wave with a wave length
corresponding to the interatomic spacing, Onsager called it the
'ghost of a vanishing vortex ring', while Chester viewed it as 'one
atom moving very fast through the sea of others'. Our arguments do
not depend on exactly what a roton is.
\bibitem{scalarproduct} The ground state and excited state wave
functions are orthogonal in the following sense: $\phi.\psi = 0 =
\int \phi\psi d\overrightarrow{R}^N$. Since $\phi$ does not have any
nodes (does not change sign), we must have that $\psi$ is positive
for half of all possible configurations $\overrightarrow{R}^N$, and
negative for the other half.
\bibitem{extend} The above arguments can be extended to
helium under pressure; we refer the reader to Ref.
[\onlinecite{wouter}] for details.
\bibitem{higherpres} This relation [Eq. \ref{rotonenergy}] can be generalized to cover
higher pressures by including a term $\sim PV$.
\bibitem{explain} The difference in specific heats is proportional to the
thermal expansion coefficient $\alpha$ as $c_p -c_v \sim \alpha^2$.
At low temperatures, $\alpha$ = 0 because the density of helium is
determined by the large zero-point-motion of the particles; reducing
the temperature will not lead to any increase in density.
\bibitem{more} The inverse of this relation [$p^*_{roton}=2\pi/d$]
gives an accurate prediction for the position of the roton minimum
for all pressures which is known to vary as $\sim n^{1/3}$, see ref.
\cite{sound}.
\end{thebibliography}
\end{document}